\documentclass[11pt,twoside]{article}

\usepackage{asp2004}
\usepackage{epsf}

\begin{document}
\title{Preliminary Results from Detector-Based Throughput Calibration of the CTIO Mosaic Imager and Blanco Telescope Using a Tunable Laser}   
\author{Christopher W. Stubbs, Sara K. Slater, Yorke J. Brown, Daniel Sherman}  
\affil{Department of Physics\\
Harvard-Smithsonian Center for Astrophysics\\
Harvard University\\
17 Oxford Street\\
Cambridge\\
 MA 02138\\
USA\\}    

\author{R. Chris Smith, Nicholas Suntzeff, Abi Saha}
\affil{Cerro Tololo Interamerican Observatory\\
NOAO\\
Casilla 603\\
La Serena\\
Chile\\
~\\}

\author{John L. Tonry, Joseph Masiero, Stephen Rodney}
\affil{Institute for Astronomy\\
University of Hawaii\\
2680 Woodlawn Drive\\
Honolulu\\
 HI, 96822\\
USA\\
}

\begin{abstract} 
We describe the scientific motivation for achieving photometric precision and
accuracy below the 1\% level, and we present a calibration philosophy 
based on using calibrated {\it detectors} rather than 
celestial {\it sources} as the fundamental 
metrology reference. A description of the apparatus and methodology is presented, as well as preliminary measurements of relative
system throughput vs.~wavelength for the Mosaic imager at the CTIO Blanco 4m telescope.
We measure the throughput of the optics, filter, and detector by comparing
the flux seen by the instrument to that seen by a precisely calibrated monitor 
photodiode,  using a tunable laser as the illumination source.  
This allows us to measure the transmission properties of the system, passband 
by passband, with full pupil illumination of the entire optical train. 
These preliminary
results are sufficiently promising that we intend to further pursue this technique, particularly
for next-generation survey projects such as PanSTARRS and LSST. 
\end{abstract}


\section{Introduction and Motivation}

The challenge of photometry is to extract knowledge of the location and flux distribution of 
astronomical sources, based on measurements of the 2 dimensional distribution of detected photons
in a focal plane. Each pixel $i$ in the detector array sees a signal $S_i$ given by

\begin{equation}
S_i = \sum_{sources~j} \int \Phi_j(\lambda) R_i(\lambda) T(\lambda)  A_i ~ d\lambda,
\label{eq:psignal}
\end{equation}

\noindent
where the sum is taken over all sources (including the sky) that contribute to the flux in the 
pixel, $\Phi_j(\lambda)$ is the photon spectrum for source $j$, $R_i(\lambda)$ is the 
throughput of the pixel, including the transmission of the optics and the pixel's
quantum efficiency,  $T(\lambda)$ is the optical transmission of the atmosphere, and 
$A_i$ is the effective aperture of the system for pixel $i$, essentially the wavelength-independent
part of the instrumental response.  

Traditional flat-fielding techniques attempt to extract knowledge of $\Phi(\lambda)$ from 
this array of sums of integrals by first dividing the flux in each pixel by a scalar number, $F_i^B$,
the ``flat field'' for that pixel in a passband $B$. It is however clear from the above equation that this is not
arithmetically correct if multiple sources with different photon spectra are contributing to the 
flux in the pixel. Furthermore, airmass corrections typically assume that the atmospheric transmission 
for a given band $B$ depends only on the secant of the zenith angle, ignoring any wavelength dependence across the 
passband. 
Astronomical instruments are currently calibrated, in practice, using celestial calibrators. 
At present the most popular photometric system is based on our knowledge 
\citep{Hayes75} \cite{Hayes75b} \cite{Castelli94} of Vega. Our observational information about the
 spectrum of Vega
is of course fundamentally based upon the terrestrial blackbody sources against which 
Vega was itself calibrated, by ground-based measurements. 
Present work, described elsewhere in this volume, uses 
a combination of photospheric modeling and observation to construct a synthetic spectra
of spectrophotometric calibration objects. Since we don't know the distances or radii of the
sources well enough to determine what their apparent fluxes should be, there is an overall
multiplicative ambiguity that is (for Vega-based magnitude systems) tied to a monochromatic
flux of Vega. This in turn links all celestial calibrators to the terrestrial blackbody 
sources that were used in the calibration of Vega. 

Considerable careful effort has been expended on building a network of spectrophotometric 
stars across the sky \cite{Oke83} \cite{Colina94}  \cite{Hamuy92}  \cite{Hamuy94}  \cite{Bohlin96} 
\cite{Megessier95}. These stars, in conjunction with the 
network of secondary photometric standards \cite{Landolt92}, form the basis for determining 
the instrumental throughput of different instruments, and for making the photometric transformations
into a ``standard'' system  \cite{Fukugita96}. 

The current state of the art with this approach produces flux measurements 
with fractional uncertainties at the few percent level \cite{SDSS04}  \cite{Bessell99}  \cite{Bessell05}. Considerable further progress along these lines is described in other contributions to this 
conference.

A variety of forefront scientific issues motivate breaking through the 1\% barrier 
that has long been a limitation in the precision of ground-based photometry. 
One example, which has in large part motivated the work described here, 
is the Dark Energy problem. Type Ia supernovae are a powerful probe of the 
history of cosmic expansion \cite{highz} \cite{scp}, and they provide strong evidence for 
the existence of the Dark Energy. As we move from the detection to the characterization 
of Dark Energy, the challenge is detecting subtle signals in the Hubble diagram. 
In particular, measuring the equation of state parameter $w=P/\rho$ of the Dark Energy
requires confidence in photometry at or below the 1\% level. 

The fundamental measurement in supernova cosmology is a determination of apparent 
brightness vs.~redshift. In order to measure brightness in the same spectral region for 
each supernova, we must shift to redder passbands for supernovae at increasing 
redshifts.  Any systematic miscalibration of the photometric zeropoints in 
different passbands would produce a corresponding systematic distortion in the
Hubble diagram. Furthermore, a detailed knowledge of the instrumental response
function is required, in order to properly account for the effects of the redshifted SN spectrum
 as seen in the passbands used (i.e.~ to perform``K corrections'' \cite{Peacock98}).

An alternative approach, which we intend to pursue, is to {\it measure} the quantities
$R_i(\lambda)$, each pixel's response vs.~wavelength, and $T(\lambda)$, the 
optical transmission of the atmosphere for each image.   We have described the
formalism of this approach elsewhere (Stubbs \& Tonry 2006) and we encourage the reader to 
consult that paper in conjunction with this one, which describes a preliminary implementation of this
technique. In Stubbs \& Tonry (2006) we point out  (for a major wide field survey which 
will measure many of currrent photometric standards as a matter of course) the merits of 
reporting all photometric survey results in the 
``natural'' photometric system of the survey, rather than inflicting the systematic errors 
introduced by transferring measurements into a ``standard'' system. 

In particular we intend to fully exploit the fact that we can obtain detectors whose response 
vs.~ wavelength can be far better characterized \cite{Larason98} than the spectrum of a calibration source. 

The measurement of atmospheric transmission is another essential ingredient in this 
approach, which is a topic of current research \cite{Adelman96} \cite{Granett05}, but we 
will not address that issue here. 

\section{Throughput Measurements of the CTIO Mosaic Imager and the Blanco 4m Telescope}

The preliminary results described here were obtained in Dec 2005, in a run that was 
undertaken to test the feasibility of the approach. 

\begin{figure}[!ht]
\begin{center}
\plotone{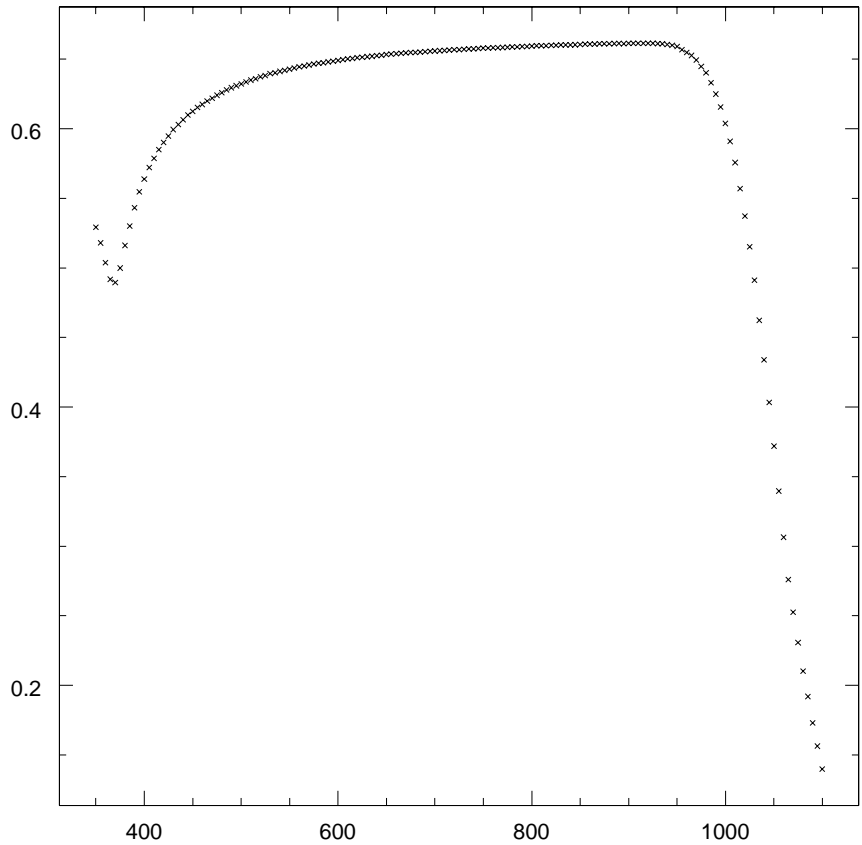}
\caption{Quantum Efficiency Curve for NIST Photodiode. This plot shows the photon detection 
efficiency vs. wavelength for a Hamamatsu 2281 photodiode. The curve is dominated by the 
index of refraction mismatch between Silicon and air; the internal QE is essentially unity for 
most of the spectral regime of interest. The lack of structure and the fact that the detector response
can be calibrated at the 10$^{-4}$ level makes these devices attractive as fundamental 
metrology standards. }
\label{fig:NIST}
\end{center}
\end{figure}

Our system uses the well-characterized detection efficiency of a NIST-calibrated
photodiode as the ``laboratory standard'' against which we calibrate the apparatus. The quantum efficiency
(QE) of the device as a function of wavelength is shown in Figure \ref{fig:NIST}.  

\begin{figure}[!ht]
\begin{center}
\plotone{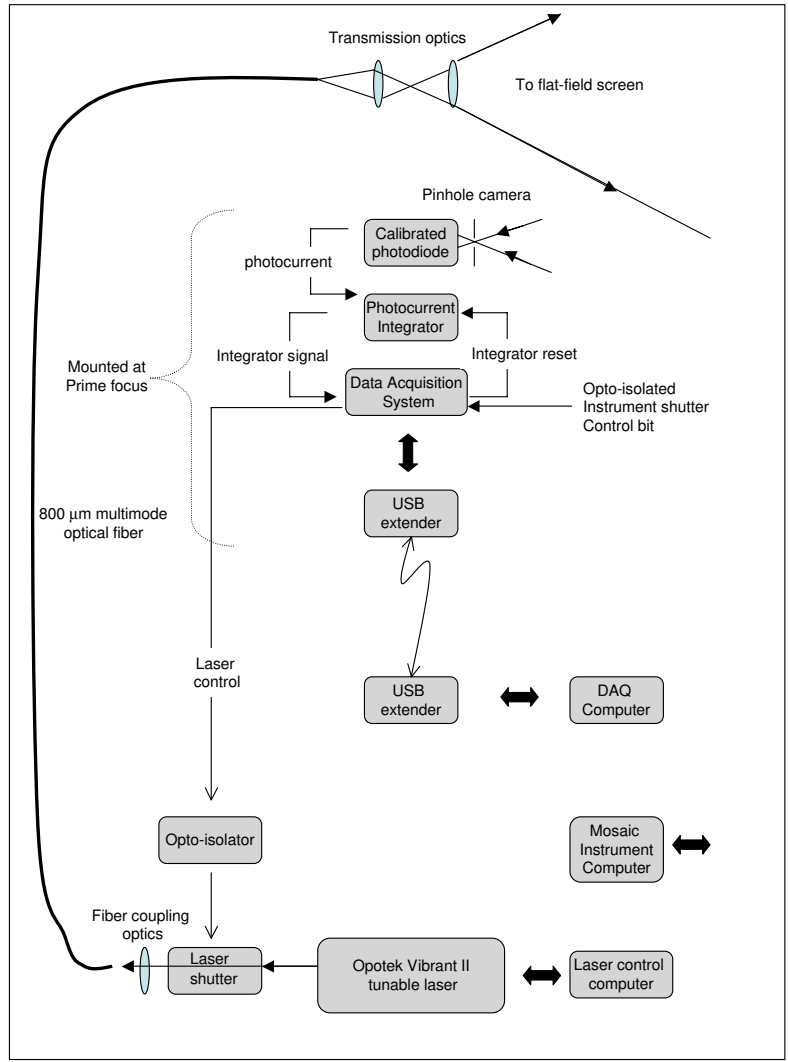}
\caption{Schematic diagram of calibration system configuration. Monochromatic light from the tunable laser is projected onto the full-aperture flat-field screen. The calibrated photodiode is used to 
monitor the total laser light delivered to the input pupil of the 4m Blanco telescope, using 
timing information derived from the Mosaic camera's shutter bit.}
\label{fig:system_diagram} 
\end{center}
\end{figure}

An overall conceptual diagram of the arrangement of the apparatus is shown in Figure \ref{fig:system_diagram}.  
We project light from a tunable laser onto the flat-field screen in the dome. We measure the 
flux reflected from the screen, incident on the telescope pupil, with a calibrated photodiode. We
then compare the flux detected by the instrument to the incident flux, as measured by the 
photodiode. Performing this measurement at a succession of wavelengths allows us to determine system throughput as a function of wavelength, using the calibrated photodiode as the 
fundamental reference. 

\subsection{The Blanco Telescope and the Mosaic II Prime Focus Imager}

We define the imager to include the detectors, dewar window and filters. The primary mirror and the 
prime focus corrector optics are considered as part of the telescope. 

\subsubsection{The Blanco Telescope}

Long a mainstay of Southern hemisphere astronomy, the Blanco telescope uses a 4m 
hyperbolic primary mirror to deliver light to (in our configuration) the prime focus. 
The $f$/2.5 beam is presented to a triplet corrector that contains an atmospheric dispersion 
compensator (ADC). 
When feeding prime focus, the telescope optics comprise one reflection from the Aluminium-coated 
primary mirror, plus four compound optical elements (the 3 corrector lenses plus the ADC). 

For the program described here we rotated the ADC to its ``neutral'' position, which 
corresponds to no correction for atmospheric dispersion. 

\subsubsection{The Mosaic II Imager}

The Mosaic II imager comprises a total of eight back-illuminated 2048 x 4096 pixel CCDs, 
with 15~$\mu$m pixels
that subtend 0.26 arcsec on a side. The total field of view of the system is 0.6 x 0.6 degrees.  Each detector is 
typically read out from two amplifiers, producing a multi-extension FITS file with 16 layers (two amplifiers for each of eight detectors). The CCD output stage and 
Arcon characteristics limit the pixel read rate to about 50 Kpix/sec, resulting in a readout time of 
120 sec for an unbinned image. This readout time dominates our overall throughput in 
acquiring a stack of calibration images, since our typical flat-field exposure time is 10 sec. 

The instrument has a filter changing mechanism that contains up to 14 large-format filters. 
Some of these are colored glass composites, and others are interference filters.  We also 
have the ability to use a blank glass ``dummy'' filter that retains the optical prescription 
by inserting a piece of fused silica glass in place of a filter. 

While there is some variation across the detectors in the Moasic array, typical 
characteristics are listed in Table \ref{table:Mosaic}. Details are available from the online documentation
that is maintained by CTIO. Dark current is negligible. 

\begin{center}
\begin{table}[htdp]
\begin{center}
\caption{Typical Mosaic Imager Characteristics}
\begin{tabular}{|c|c|}
\hline
Parameter & Typical Value\\
\hline
Read Noise & 6 electrons\\
Gain & 2 electrons/ADU\\
Saturation (min) & 60,000 electrons\\
Crosstalk between amplifiers & 10$^{-3}$\\ 
\hline
\end{tabular}
\end{center}
\label{table:Mosaic}
\end{table}%
\end{center}

For the calibration data described here, we were careful to keep the signal levels 
within the linear regime for the most restrictive of the detectors.

\subsection{Calibration Apparatus}

The tunable laser was located in a room within the 4m dome, to shield it from 
temperature changes and from dust. An 800 micron diameter 
multimode optical fiber with low OH content, clad in semirigid 
conduit, carried the laser light up
through the telescope structure and onto an optical bench assembly that was mounted on the 
top of the prime focus cage. This optical bench held the optics for illuminating 
the flat-field screen, the calibration diode, the photocurrent integration 
electronics, and our A/D converter. A USB extender carried digital signals between this 
location and the data acquisition and control computer that was located near the laser. 
Computers for controlling the laser and the Mosaic instrument were also placed 
at this location. 

We illuminated the full area of the flat-field screen by sending light that emerged from the 
fiber optic through a pair of lenses. 
Each unvignetted element of area $dA$ on the flat-field screen illuminates every pixel of the imager, 
since the optical system maps angles at the pupil to position in the focal plane. The camera 
sees all rays that emerge within a cone of opening angle $\sqrt{2} \times 0.3 \deg = 0.42 \deg$. 
If the screen emission is approximated as a Lambertian, then the most extreme ray has an 
intensity relative to the central ray of  $I(edge)/I(center) = cos(0.42\deg) = 0.9999$. 
Although we would have ideally mapped out the Bi-Directional Reflectance Function (BDRF) 
of the flat-field screen as a function of wavelength, this was impractical as the flat-field screen surface
was at an inaccessible height in the dome. 

\subsection{Flat-field Screen Illumination.}

We used a Vibrant II tunable laser (manufactured by Opotek of Carlsbad CA) as a tunable source of monochromatic light. The laser uses a 20 Hz Nd:YAG laser at 1.064 microns as the source
of photons. This light passes through a pair of non-linear optical crystals that upconvert the light into photons at $\lambda$=355 nm. These UV photons are then run through an optical parametric oscillator (OPO) that downconverts each UV photon into a pair of photons, conserving both energy and momentum. 
The orientation of the OPO crystal relative to the incident beam can be adjusted so as to select
for a specific wavelength of interest. The wavelength of the output beam can thereby be tuned over a wavelength range of 
400~nm to 2~$\mu$m. 

From the photon pair produced, the upper or lower frequency beam is selected by exploiting the fact that these two beams have orthogonal linear polarizations. 
Our light source produces 5 nsec wide pulses at 20 Hz, with typical energies of 10-50 mJ 
per pulse. We measured a power of 10 mW
going into the fiber at $\lambda$=800 nm when we adjusted the power to give 30,000 ADUs
of typical signal in the Mosaic pixels. There is a ``degeneracy point'' is in the OPO system at $\lambda$=710~nm where the OPO output is not well behaved, but this is a very narrow spectral avoidance 
region and did not pose a problem. 
 
We also use optical filters to ensure that there is no contamination 
from the partner photon, or from the UV beam.  
We measured the spectral contamination from the partner photon to be 
less than one part in 10$^3$.
 
The light intensity from the laser is adjustable in two ways. The Nd:YAG laser has an variable
time delay between the flashlamp pulse and the Q-switch driven dump of the laser cavity. 
Changing this time delay varies the intensity of the light emitted by the Nd:YAG laser. In addition 
we installed an actuated polarizer on the output of the OPO stage, that can be used to 
modify the intensity of the monochromatic light downstream of the OPO. We used a combination of
these to generate a desired intensity of light from the tunable laser. Since the conversion efficiency of the 
OPO system does depend on wavelength, we found the ability to adjust the output intensity to be an important feature of the illumination system. 

The light from the tunable laser was focussed onto a multimode optical fiber which was run through the telescope structure and up to the prime focus cage. The output end of the fiber 
was attached to an optical bench that was mounted on the upper end of the prime focus cage. 
The 800 micron output tip of the fiber was then imaged onto the flat-field screen, making a spot
that filled roughly 75\% of the screen area.

Although the light emitted from the laser is polarized, as the light is transmitted through the 
optical fiber its polarization becomes randomized. The coherence length of the pulsed laser
light is sufficiently short that speckle effects were unobservable.  

\subsection{Monitoring light at the input pupil with Calibrated Photodiode}

We configured a NIST-calibrated Silicon photodiode  as a pinhole camera. By avoiding 
any imaging optics we avoid introducing any unwanted wavelength dependence in the calibration
signal chain. 

Ideally the calibration diode would monitor the entire emitting area of the flat-field screen 
over the angular range seen by the camera system, but this was not practical:  we have a standoff distance of about 4~m between the flat-field screen and the calibration detector. 
We had the choice between configuring the diode to monitor
the entire screen area over a wider angle, or narrowing the calibration diode's field of view
and monitoring a portion of the screen area. We elected to configure the diode to 
monitor the full illuminated region of the screen, and adjusted the pinhole spacing to 
accomplish this.

\subsection{Electronics and Instrument Interface}

The system elements include a data acquisition module which 
was connected to a computer through a USB extension module. This allowed 
our main system components (except for the laser) to be mounted on the top end ring of the telescope. 

Taking a dome flat image with the Mosaic camera, and more specifically the opening of the Mosaic shutter, initiated a data taking sequence. The shutter control bit was therefore 
passed through an opto-isolation stage and then monitored by our data acquisition module. 

The photodiode output was sent into an integrator circuit that we used to monitor the integrated
dose of light received by the screen. This integrative approach was important in 
minimizing any deleterious effects due to pulse-to-pulse intensity variations in the laser output. 
The integrator output was connected to one of the analog inputs on the data acquisition module. 
The integrator was reset on command using one of the digital outputs on the data acquisition module. 

We used one laptop computer to control the tunable laser, a second to run the Mosaic camera, 
and a third one to communicate with the data acquisition system. 

\subsection{Data Acquisition Sequence}

We rotated the Atmospheric Dispersion Corrector (ADC) to the neutral position, 
and pointed the telescope to the nominal location of the flat-field screen. We then 
acquired calibration images according to the following prescription, iterating 
through wavelength:

\begin{enumerate}

\item{}  We selected the filter of 
interest. 

\item{} We adjusted the laser wavelength to the desired value, changing the polarizer and 
wavelength ``cleanup'' filter as appropriate. 

\item{} We adjusted the flashlamp to Q-switch delay and the attentuator setting on the Opotek laser 
while monitoring the light intensity with the monitor photodiode, to provide uniform intensity. This 
also ensured that we delivered the full light dose during the 10 sec interval while the Mosaic shutter was open. 

\item{} We took a ``laser on'' dome flat exposure with the Mosaic imager, with a typical 
exposure time of 10 sec. 
As described above 
our electronics monitored the shutter status bit and once the Mosaic shutter is fully open first 
reset the photocurrent integrator and, after a half second delay, we open the laser shutter and allow the laser light to strike the flat-field screen. The integrated signal from the photodiode is monitored by
our data acquisition system, and once a calibrated dose of photons is delivered to the input pupil
of the telescope we close the laser shutter. We are careful to ensure that the laser shutter closes
before the Mosaic shutter. This eliminates potential systematic effects from the Mosaic shutter. 
We stored both the FITS file from the Mosaic imager as well as the time history of the 
integral of the flux seen by the photodiode. 

\item{} We then took a ``laser off'' exposure with the Mosaic imager, in order to measure the ambient 
light contamination in our flats. In our processing we subtract these adjacent ``laser-off'' images
in order to use only the laser light contributions to the flats. We again store the associated time history of the photodiode flux.  

\item{} We then changed to the next wavelength of interest, and iterated the procedure. 

\end{enumerate}

\subsection{Representative Raw Data}

\begin{figure}[!ht]
\begin{center}
\plottwo{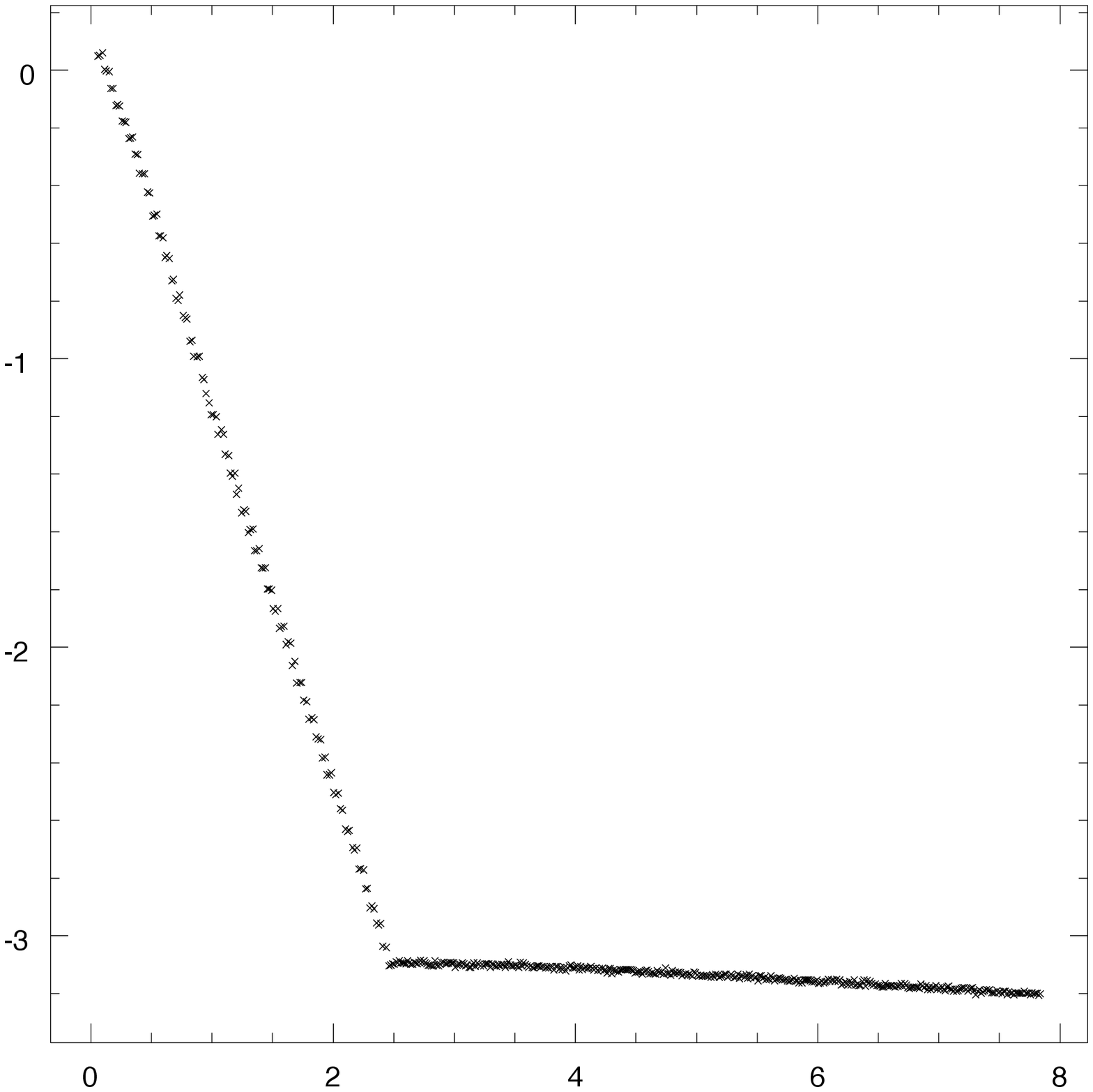}{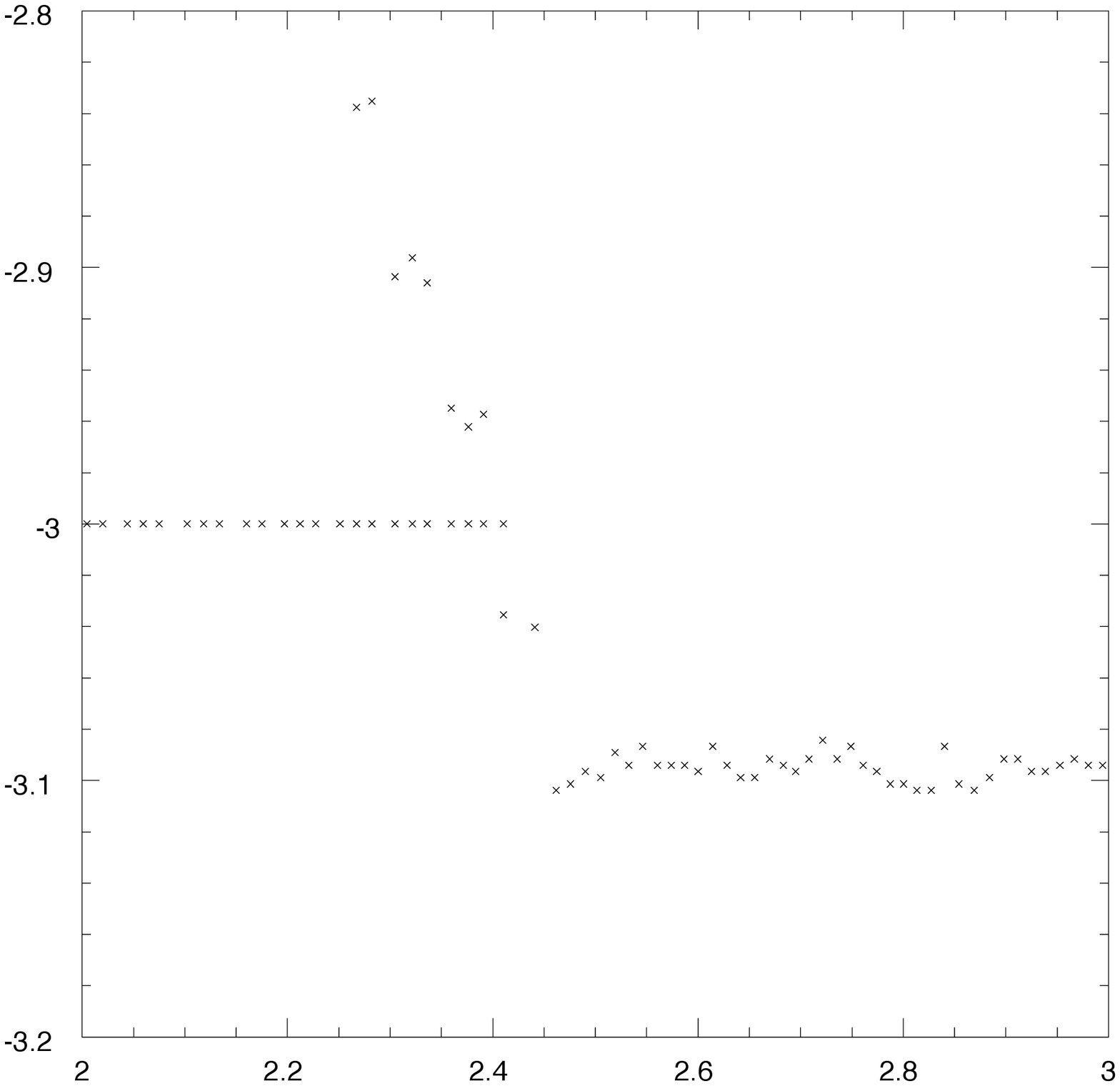}
\caption{Integrated signal from photodiode. These plots show the integrated photodiode signal 
in arbitrary units vs.~time (in sec). More light drives this signal more negative. The left hand 
panel shows the integrated light at the photodiode over an entire 
image, where the steep region is with the laser on and the flatter region is after laser shutoff. 
The laser light clearly dominates over the ambient light in the dome, even in the daytime.
We did linear fits to distinguish between the contributions from ambient and laser light.  
The right hand panel is an expansion of the time around laser shutoff. The individual laser pulses are clearly visible. The horizontal line at the -3 level is the laser shutter command bit status. }
\label{fig:photodiode} 
\end{center}
\end{figure}

Figure \ref{fig:photodiode} shows a typical measurement of the integrated light intensity seen by the photodiode. The 
plot shows the ambient and laser light contributions. The
laser intensity is at least ten times that of the ambient light in the dome.  

\begin{figure}[!ht]
\begin{center}
\plottwo{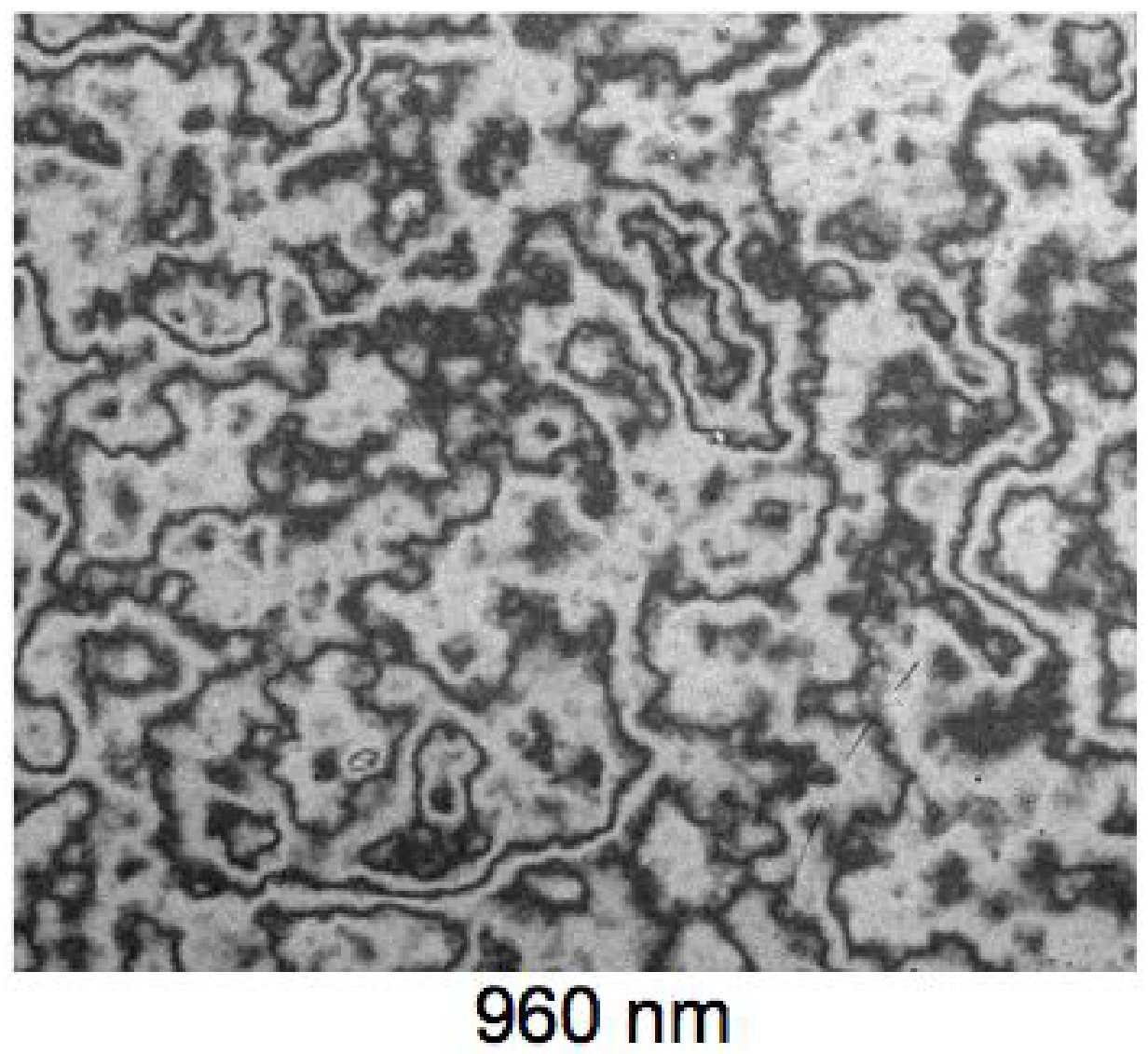}{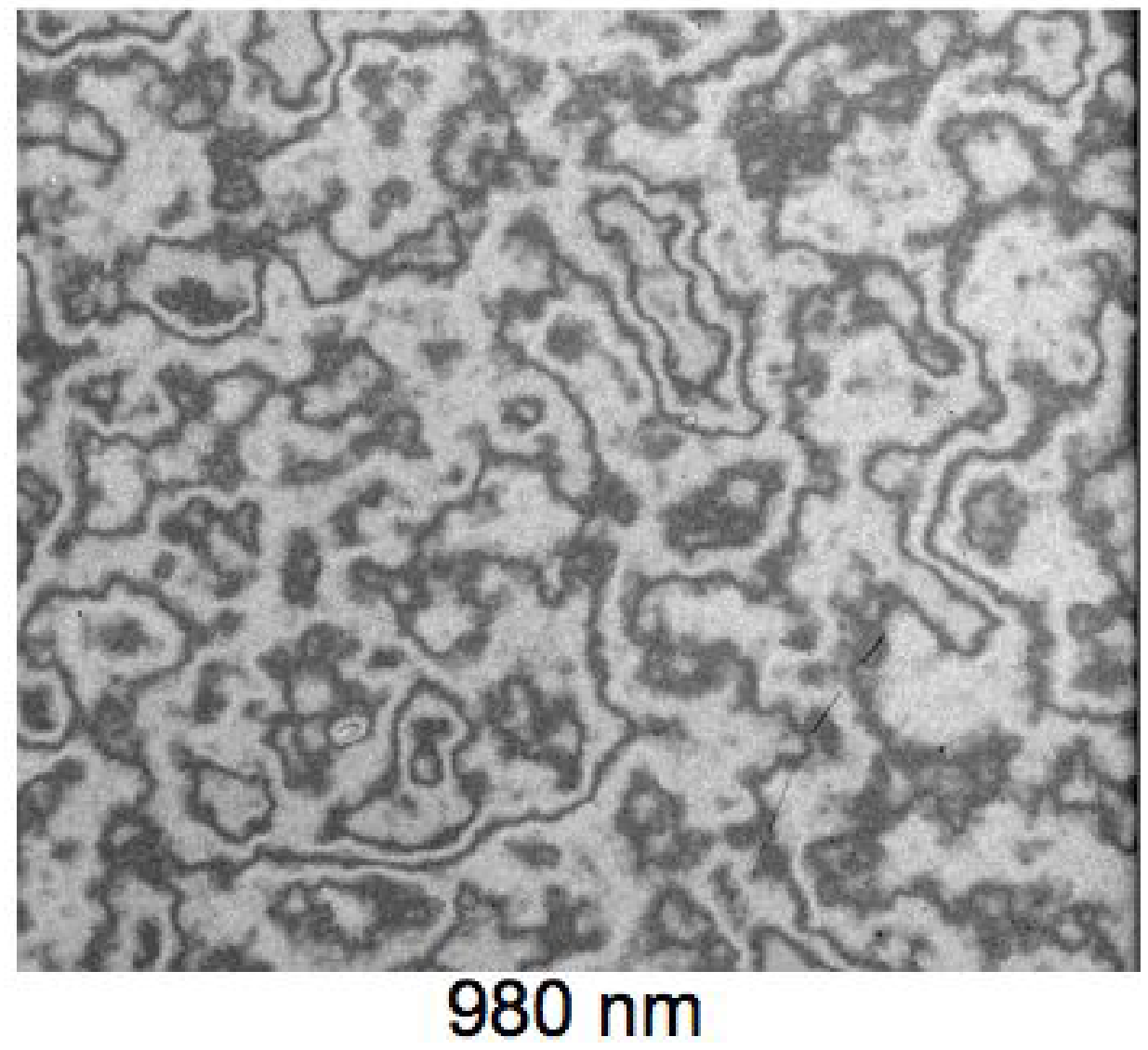}
\caption{Fringing Measured by Monochromatic Laser Illumination. These two panels compare the 
fringing pattern at $\lambda$=960~nm and $\lambda$=980~nm. Note the variation in phase of the
fringing pattern between these two wavelengths. Data such as these should allow us to 
construct high accuracy fringe frames for flatfielding, in conjunction with sky emission spectra.}
\label{fig:fringing}
\end{center}
\end{figure}

An amusing example of a pair of images from the Mosaic instrument is shown in Figure \ref{fig:fringing}, 
which clearly shows the spatial modulation of device QE that is responsible for ``fringing" 
due to night sky emission lines.  At longer wavelengths the fractional variation in a pixel's  QE over wavelength can be as 
large as 10\%. 

Figure \ref{fig:fringing} illustrates that spatial variation in QE not only produces additive problems 
arising from bright sky lines,  
which are traditionally removed through fringe corrections, but also introduce subtle 
variations in effect wavelength response across the array which are manifested as spatial variations 
in the effective passband. It is awkward that the spatial scale of the fringing variations is matched
to the size of a typical PSF. For astronomical sources with spectra that are not flat, this spatial QE variation is not properly corrected for by using broadband flats. For example the 
photometry reported for an emission line object could vary by as much as 10\%, depending on where
on the array the object is located. Furthermore, this effect suggests that narrowband imaging
is more compromised than broadband imaging, where many cycles of fringing are averaged
over, across the optical passband. 

\section{Data Processing}

\subsection{Calibration Signal Processing}

We stored the calibration photocurrent signal before, during and after the laser shutter was opened. 
As shown in Figure \ref{fig:photodiode} we did linear fits to determine the ambient light intensity, 
$\Phi_{ambient}^{diode}$, in ADUs per second. We subtracted this from each data point and 
used the linear fit to the integrated photocurrent to determine the total integrated
laser light dose $I_{laser}^{diode}(\lambda)$ (in ADUs) for each measurement.  Dividing this 
value by the diode QE provides us with a number that is proportional to the 
number of photons delivered to the input pupil;
 $N_{laser}(\lambda)={{I_{laser}^{diode}(\lambda)}  \over{ QE(\lambda)}}$.
 
\subsection{Processing of the CCD Images}

The images, both laser-on and laser-off, were corrected for bias levels using the 
pixel overscan regions. 
The laser-off images showed only a very slow variation over time, in agreement with the 
ambient light levels reported by the calibration diode. The difference between 
laser-on and laser-off images produced an image array $F(i,j,\lambda)$ 
that contains a measurement of throughput for each pixel $(i,j)$ at the 
wavelength $\lambda$. 

\subsection{Determination of System Throughput}

\begin{figure}[!ht]
\begin{center}
\plottwo{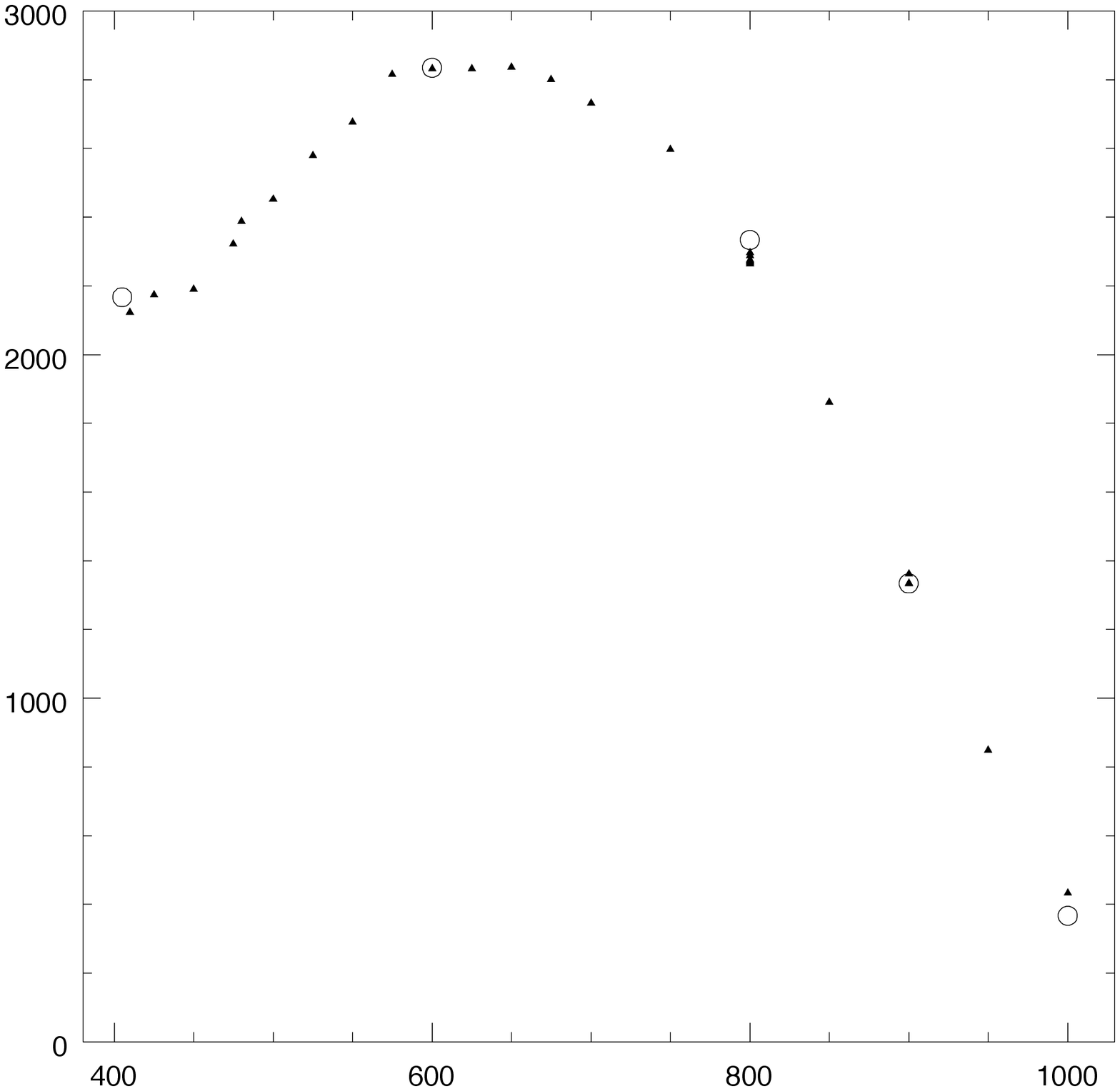}{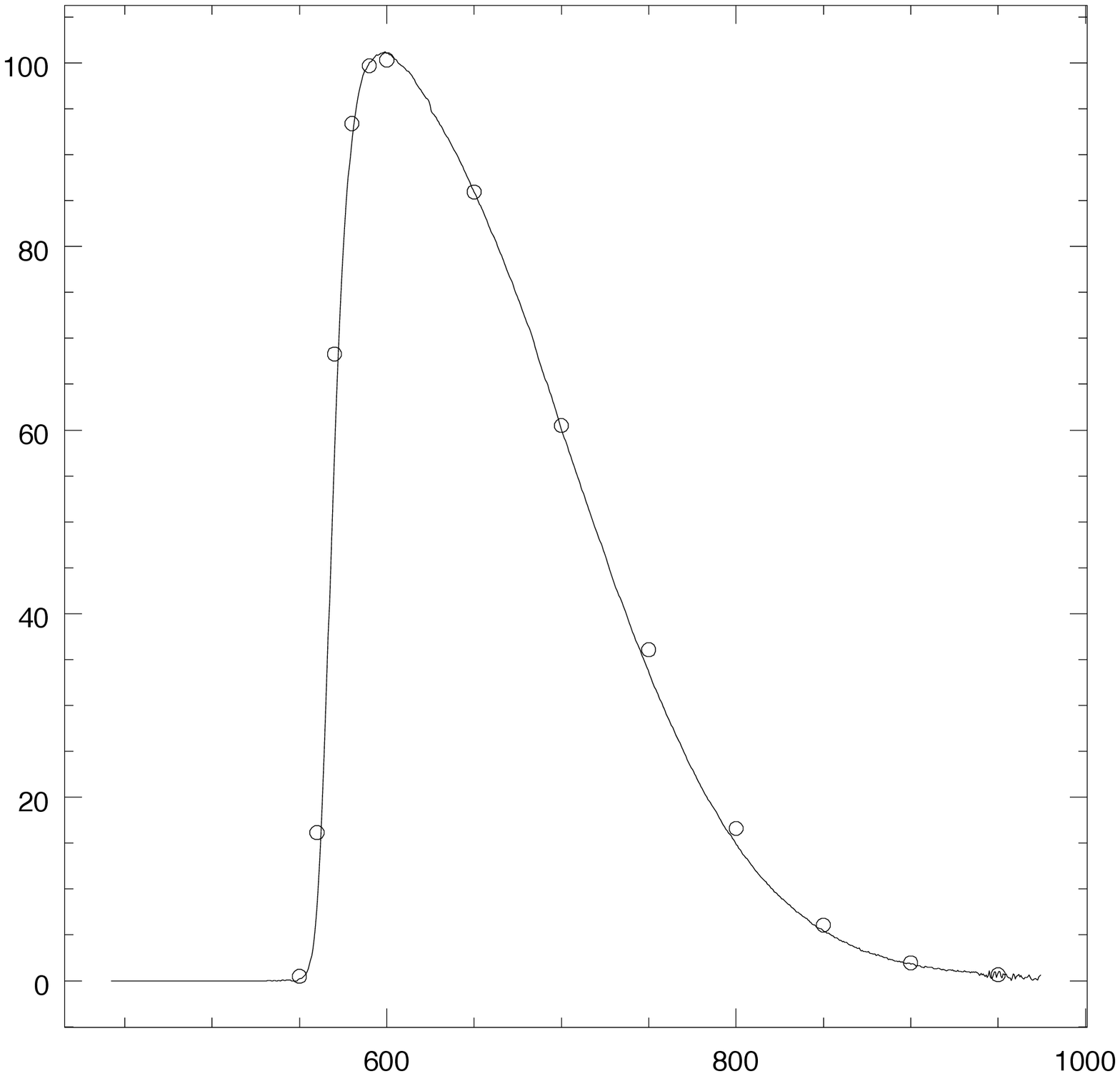}
\caption{Throughput with Fused Silica Blank Filter and R band filter sanity check. 
The left panel shows relative system 
throughput (in arbitrary units) vs.~wavelength in nm, 
obtained with a fused silica blank in place of a filter. The solid points are the data we
obtained and the open circles are the vendor's room temperature measurements of the 
detector QE. We multiplicatively scaled our data by single overall normalization in order to 
drive agreement at $\lambda$=600 nm. The right panel shows a consistency check of R band filter throughput. We took throughput data with the R
band filter, made from a stack of Schott glasses, and 
divided it by the blank glass throughput curve shown in 
left panel. This ratio should reproduce the throughput of the R band filter alone. 
The right hand figure shows (as circles) our data while the solid line is the the transmission curve
for this  filter from the documentation maintained at CTIO. The good agreement 
shows that this approach of in-situ, full aperture throughput determination is promising. }
\label{fig:blank}
\end{center}
\end{figure}

For each wavelength we took the ratio $T(i,j,\lambda) = {{F(i,j,\lambda) } \over {N_{laser}(\lambda)}}$
to determine each pixel's sensitivity at the wavelength $\lambda$.  We took a 100 x 100 pixel region 
in one of the amplifiers to generate the representative results shown here.  Figure \ref{fig:blank}
shows throughput vs.~wavelength when the filter is replaced by a fused silica blank. We also show the vendor's QE (at room temperature) for a device that is representative of those mounted in the 
Mosaic camera. 

We took multiple data sets at $\lambda$=800 nm in order to (1) determine repeatability and 
(2) ascertain whether the results depended upon the total light dose. Changing the
light dose by a factor of two had no measurable effect. The throughput measurements are
reproducible to better than 1\%.  

\section{Preliminary Conclusions, and Next Steps}

We consider the preliminary results we obtained to be sufficiently promising to proceed with 
this calibration scheme. We note that the alternative approach we are pursuing does not in any
way compromise or collide with reducing survey data with standard techniques. In fact comparing
the results with the two methods will provide useful constraints on systematic effects. 

Our near term goal is to obtain a full set of throughput calibration data for the various filters 
on the Mosaic instrument, by the end of 2006. We also will

\begin{itemize}

\item{} Develop and refine atmospheric transmission measurement techniques, 

\item{} Develop and refine self-luminous flat field screens, as described in the companion
paper by Brown et al in these proceedings, 

\item{} Continue to work towards implementing these techniques for both PanSTARRS
and LSST.

\end{itemize}

\acknowledgements 

We are grateful to the US National Science Foundation, under grants AST-0443378 and 
AST-0507475, for their
support of this work in the context of the ESSENCE supernova survey. We are also 
grateful to the LSST Corporation, Harvard University, and the US Department of 
Energy Office of Science
for their support of the continuing development of these
techniques. The work described here would not have been possible without the 
skill and support of the technical staff at CTIO. We are also very grateful to Dr.~Eli Margalith
of Opotek for his support and assistance in this endeavor. 


\end{document}